\documentclass[sigconf]{acmart} 

\usepackage{booktabs}
\usepackage{graphicx, epstopdf}
\usepackage{subfigure}
\usepackage{url}
\usepackage{multirow}
\usepackage{algorithm}
\usepackage{algorithmicx}
\usepackage{algcompatible}
\usepackage{balance}
\usepackage{arydshln}
\usepackage[normalem]{ulem}

\renewcommand*{\textrightarrow}{$\rightarrow$}

\copyrightyear{2024}
\acmYear{2024}
\setcopyright{acmlicensed}\acmConference[SIGIR '24]{Proceedings of the 47th International ACM SIGIR Conference on Research and Development in Information Retrieval}{July 14--18, 2024}{Washington, DC, USA}
\acmBooktitle{Proceedings of the 47th International ACM SIGIR Conference on Research and Development in Information Retrieval (SIGIR '24), July 14--18, 2024, Washington, DC, USA}
\acmDOI{10.1145/3626772.3657941}
\acmISBN{979-8-4007-0431-4/24/07}

\begin{document}

\title{FedUD: Exploiting Unaligned Data for Cross-Platform Federated Click-Through Rate Prediction}

\author{Wentao Ouyang}
\affiliation{%
  \institution{Alibaba Group}
  \city{Beijing}
   \country{China}
}
\email{maiwei.oywt@alibaba-inc.com}

\author{Rui Dong}
\affiliation{%
  \institution{Alibaba Group}
  \city{Beijing}
  \country{China}
}
\email{kailu.dr@alibaba-inc.com}

\author{Ri Tao}
\affiliation{%
  \institution{Alibaba Group}
  \city{Beijing}
  \country{China}
}
\email{taori.tr@alibaba-inc.com}

\author{Xiangzheng Liu}
\affiliation{%
  \institution{Alibaba Group}
  \city{Beijing}
  \country{China}
}
\email{xiangzheng.lxz@alibaba-inc.com}

\begin{abstract}
Click-through rate (CTR) prediction plays an important role in online advertising platforms. Most existing methods use data from the advertising platform itself for CTR prediction. As user behaviors also exist on many other platforms, e.g., media platforms, it is beneficial to further exploit such complementary information for better modeling user interest and for improving CTR prediction performance. However, due to privacy concerns, data from different platforms cannot be uploaded to a server for centralized model training. Vertical federated learning (VFL) provides a possible solution which is able to keep the raw data on respective participating parties and learn a collaborative model in a privacy-preserving way. However, traditional VFL methods only utilize aligned data with common keys across parties, which strongly restricts their application scope. In this paper, we propose FedUD, which is able to exploit unaligned data, in addition to aligned data, for more accurate federated CTR prediction. FedUD contains two steps. In the first step, FedUD utilizes aligned data across parties like traditional VFL, but it additionally includes a knowledge distillation module. This module distills useful knowledge from the guest party's high-level representations and guides the learning of a representation transfer network. In the second step, FedUD applies the learned knowledge to enrich the representations of the host party's unaligned data such that both aligned and unaligned data can contribute to federated model training. Experiments on two real-world datasets demonstrate the superior performance of FedUD for federated CTR prediction.
\end{abstract}



\settopmatter{printacmref=true}
\renewcommand{\shorttitle}{FedUD: Exploiting Unaligned Data for Cross-Platform Federated CTR Prediction}

\maketitle

\vspace{-3.6pt}
\section{Introduction}
Click-through rate (CTR) prediction is one of the most central tasks in online advertising platforms \cite{zhou2018deep,ouyang2019deep}.
Most existing methods use data collected from the advertising platform itself for CTR prediction \cite{he2014practical,cheng2016wide,zhang2016deep,shah2017practical,wang2017deep,zhou2018deep,ouyang2019deep,song2019autoint,qin2020user,tian2023eulernet}.
As user behaviors also exist on many other platforms, e.g., media platforms, it is beneficial to further exploit such complementary information for improving CTR prediction performance \cite{wu2022fedctr,ouyang2020minet}. However, due to user privacy concerns, data from different platforms cannot be uploaded to a server for centralized model training. Vertical federated learning (VFL) provides a feasible solution which allows different participating parties to keep raw data locally but train a collaborative model in a privacy-preserving way \cite{yang2019federated}.

However, traditional VFL methods only utilize aligned data with common keys across parties, which strongly restricts their application scope. In this paper, we propose FedUD, which is able to exploit unaligned data, in addition to aligned data for federated CTR prediction. For simplicity, we consider a two-party setting, where the target \textbf{advertising platform} serves as the \textbf{host party} and another \textbf{media platform} servers as the \textbf{guest party}. FedUD contains two steps. First, FedUD utilizes aligned data across parties like traditional VFL, but it additionally includes a knowledge distillation module. Second, FedUD applies the learned knowledge to enrich the representations of the host party's unaligned data.

\begin{figure*}[!t]
\vskip -10pt
\centering
\subfigure[Step 1]{\includegraphics[width=0.40\textwidth, trim = 0 0 0 0, clip]{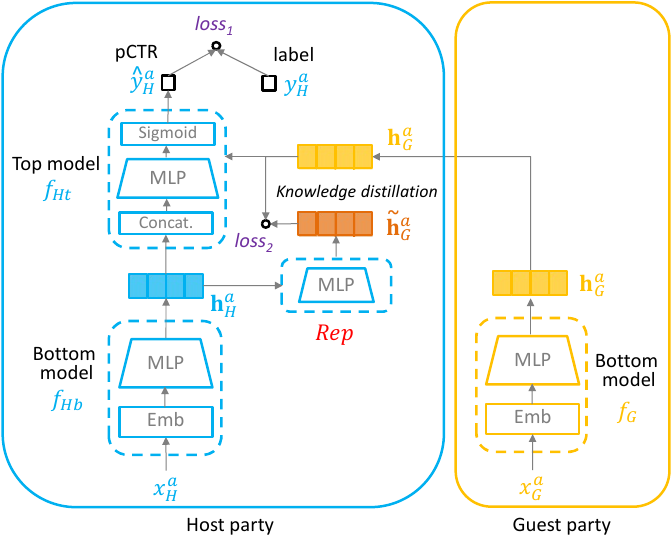}} \hskip 20pt
\subfigure[Step 2]{\includegraphics[width=0.495\textwidth, trim = 0 0 0 0, clip]{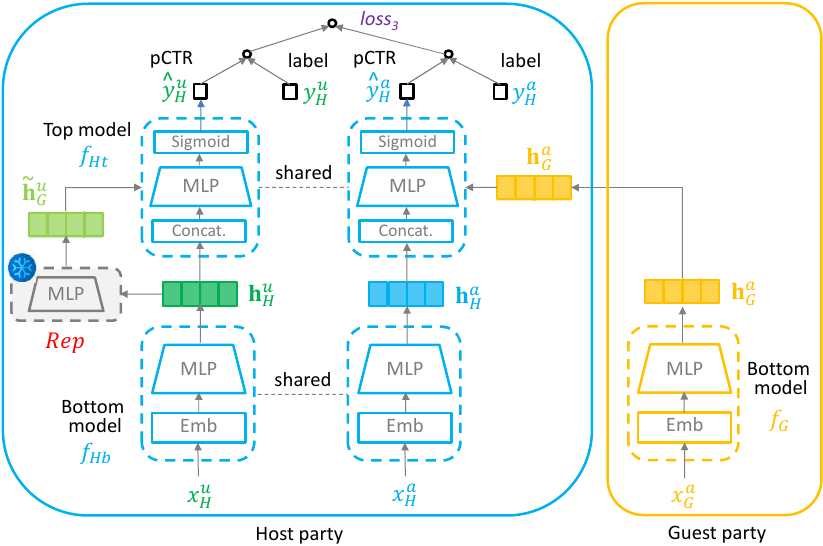}}
\vskip -12pt
\caption{Illustration of the two steps in FedUD. (a) Step 1: Federated learning using aligned data across parties $\{x_H^a, y_H^a, x_G^a\}$ with knowledge distillation. (b) Step 2: Federated learning using both aligned data across parties and unaligned data $\{x_H^u, y_H^u\}$.}
\vskip -10pt
\label{fig_model}
\end{figure*}


In summary, the main contributions of this paper are
\begin{itemize}
\item We propose FedUD for federated CTR prediction in a privacy-preserving way. FedUD is able to distill knowledge from aligned data' federated representations to enrich the host party's unaligned data' representations, such that both aligned and unaligned data can contribute to model training.
\item During inference, FedUD benefits not only aligned inference data (like traditional VFL methods), but also unaligned inference data because of its representation transfer ability.
\item We conduct experiments on two real-world datasets to evaluate the effectiveness of various methods.
\end{itemize}

\section{Problem Statement}
For ease of illustration, we consider a two-party VFL setting. The host party has labels and features, and the guest party only has additional features.
The two parties' samples have different feature spaces.
Each sample has a key. In federated CTR prediction, the sample key could be the user's \emph{hashed device ID}.
VFL needs to first perform secured key intersection, e.g., private set intersection (PSI) \cite{de2010practical, pinkas2018scalable}, to find aligned data with common keys across parties.

We denote the aligned data across parties as $\mathcal{D}^a = \{x_H^a, y_H^a, x_G^a\}$, where $\{x_H^a\}$, $\{y_H^a\}$ are the aligned samples and labels of the host party and $\{x_G^a\}$ is the aligned samples of the guest party.
Traditional VFL methods utilize aligned data to train a federated model $f(\mathcal{D}^a)$ for the host party with enriched features.

We denote the unaligned data of the host party as $\mathcal{D}^u = \{x_H^u, y_H^u\}$ where $\{x_H^u\}$, $\{y_H^u\}$ are the unaligned samples and labels.
In this paper, we aim to further exploit the useful information contained in unaligned data, in addition to aligned data, to train a federated model $f(\mathcal{D}^a, \mathcal{D}^u)$ for the host party with improved performance.

\section{Model Design}
\subsection{Step 1: Federated Learning using Aligned Data with Knowledge Distillation}
We illustrate this step in Figure \ref{fig_model}(a).
For privacy protection, each party has its own local models.
The guest party has only a local bottom model $f_G$, whose aim is to extract high-level representations based on the guest party's local data.
For simplicity, the guest party's local model $f_G$ contains an embedding layer and a multi-layer perceptron (MLP) with several fully-connected layers (ReLU nonlinear activation) \cite{nair2010rectified}.
We denote the high-level representation of the data in the guest party as
\[
\mathbf{h}_G^a = f_G(x_G^a) = MLP_G \left(\textit{Emb}_G(x_G^a) \right).
\]

The host party has a local bottom model $f_{Hb}$, whose aim is to extract high-level representations based on the host party's local data.
Similarly, $f_{Hb}$ contains an embedding layer and an MLP.
We denote the high-level representation of the data in the host party as
\[
\mathbf{h}_H^a = f_{Hb}(x_H^a) = MLP_{Hb} \left(\textit{Emb}_H(x_H^a) \right).
\]
The host party has an additional top model $f_{Ht}$, which takes the high-level representations $\mathbf{h}_H^a$ and $\mathbf{h}_G^a$ as input, and produces a CTR prediction logit as
\[
z_H^a = f_{Ht}(\mathbf{h}_H^a, \mathbf{h}_G^a) = MLP_{Ht} \left(Concat(\mathbf{h}_H^a, \mathbf{h}_G^a) \right).
\]
The host party's top model $f_{Ht}$ contains a concatenation layer and an MLP.
The predicted CTR is given by
\[
\hat{y}_H^a = Sigmoid(z_H^a).
\]

So far, we only utilize aligned data but ignores possibly useful information contained in unaligned data.
However, it is non-trivial to utilize unaligned data of the host party because these data do not have corresponding features in the guest party.

To tackle this problem, we aim to learn a representation transfer network $Rep$ where the input is the high-level representation $\mathbf{h}_H^a$ in the host party, and the output is $\widetilde{\mathbf{h}}_G^a = Rep(\mathbf{h}_H^a)$ which mimics the high-level representation $\mathbf{h}_G^a$ in the guest party in certain metric. In this way, the knowledge in $\mathbf{h}_G^a$ is distilled into $\widetilde{\mathbf{h}}_G^a$ and such knowledge guides the learning of the representation transfer network. For simplicity, this network contains an MLP.

In Step 1 of FedUD, we have two aims: 1) learn accurate high-level representations and 2) learn an accurate representation transfer network with knowledge distillation.
To achieve the first aim, we optimize the prediction loss as follows
\[
loss_1 = \frac{1}{|Y_H^a|} \sum_{y_H^a \in Y_H^a} \left[- y_H^a \log(\hat{y}_H^a) - (1 - y_H^a) \log(1- \hat{y}_H^a) \right],
\]
where $Y_H^a = \{ y_H^a\}$. 

To achieve the second aim, we optimize the mean squared error (MSE) loss between two representations as follows
\[
loss_2 = \frac{1}{|Y_H^a|} \sum_{y_H^a \in Y_H^a} \big\| \widetilde{\mathbf{h}}_G^a - \mathbf{h}_G^a \big\|^2 = \frac{1}{|Y_H^a|} \sum_{y_H^a \in Y_H^a} \big\|Rep(\mathbf{h}_H^a) -  \mathbf{h}_G^a \big\|^2.
\]

Both $loss_1$ and $loss_2$ are applied only on aligned data across parties. The overall loss in Step 1 of FedUD is given by
\[
loss_1 + \alpha \ loss_2,
\]
where $\alpha$ is a tunable balancing hyperparameter.

\subsection{Step 2: Federated Learning using Both Aligned and Unaligned Data}
We illustrate this step in Figure \ref{fig_model}(b).
For unaligned data $x_H^u$ of the host party, we can only obtain the host party's representation $\mathbf{h}_H^u$ but not the guest party's representation. This makes unaligned data useless for traditional VFL methods.

Differently, we have learned a representation transfer network $Rep$ in Step 1. Although $Rep$ is learned based on aligned data, we apply it to unaligned data to infer $\widetilde{\mathbf{h}}_G^u$ based on $\mathbf{h}_H^u$. By doing so, we can use both aligned and unaligned data for model training.

For aligned data $\mathcal{D}^a = \{x_H^a, y_H^a, x_G^a\}$, the high-level representation of the host party and that of the guest party are given by respective local bottom models
\[
\mathbf{h}_H^a = f_{Hb}(x_H^a), \ \mathbf{h}_G^a = f_G(x_G^a).
\]
The host party's top model then generates the predicted CTR $\hat{y}_H^a$ as
\[
z_H^a = f_{Ht}(\mathbf{h}_H^a, \mathbf{h}_G^a), \ \hat{y}_H^a =Sigmoid(z_H^a).
\]

For unaligned data $\mathcal{D}^u = \{x_H^u, y_H^u\}$, the high-level representation of the host party is given by its local bottom model, but the high-level representation of the guest party is given by the representation transfer network $Rep$  of the host party
\[
\mathbf{h}_H^u = f_{Hb}(x_H^u), \ \widetilde{\mathbf{h}}_G^u = Rep(\mathbf{h}_H^u).
\]
The host party's top model then generates the predicted CTR $\hat{y}_H^u$ as
\[
z_H^u = f_{Ht}(\mathbf{h}_H^u, \widetilde{\mathbf{h}}_G^u), \ \hat{y}_H^u = Sigmoid(z_H^u).
\]

We then optimize the prediction loss based on both aligned and unaligned data as
\begin{align}
loss_3 & = \frac{1}{|Y_H^a|} \sum_{y_H^a \in Y_H^a} \left[- y_H^a \log(\hat{y}_H^a) - (1 - y_H^a) \log(1- \hat{y}_H^a) \right] \nonumber\\
& + \beta \frac{1}{|Y_H^u|} \sum_{y_H^u \in Y_H^u} \left[- y_H^u \log(\hat{y}_H^u) - (1 - y_H^u) \log(1- \hat{y}_H^u) \right], \nonumber
\end{align}
where $Y_H^a = \{ y_H^a\}$, $Y_H^u = \{ y_H^u\}$ and $\beta$ is a tunable balancing hyperparameter.
When we optimize $loss_3$, we load the learned parameters of the representation transfer network $Rep$ and keep them frozen.

\subsection{Privacy}
FedUD preserves all the privacy properties of traditional VFL \cite{yang2019federated} because the guest party only sends the high-level representations $\mathbf{h}_G^a$ of aligned data to the host party and the host party only sends the corresponding partial gradients $\nabla \mathbf{h}_G^a$ to the guest party. No raw data is communicated between the two parties. Unaligned data of the host party is only utilized by the host party itself.

\section{Experiments}

\subsection{Datasets}
(1) Avazu dataset \cite{avazu-ctr-prediction}. We split it as training, validation and testing datasets each with 8 days, 1 day and 1 day of data. The host party and the guest party have 10 and 12 feature slots respectively. The sample key is hashed device ID.

(2) Industrial dataset. We split it as training, validation and testing datasets each with 7 days, 1 day and 1 day of data. The host party is a news feed advertising platform. The guest party is a media platform. The host party and the guest party have 22 and 12 feature slots respectively. The sample key is hashed device ID.

\subsection{Methods in Comparison}
\begin{itemize}
\item \textbf{DNN} \cite{cheng2016wide}. Deep Neural Network. It contains an embedding layer, several fully connected layers and an output layer.
\item \textbf{Wide\&Deep} \cite{cheng2016wide}. It combines logistic regression and DNN.
\item \textbf{DeepFM} \cite{guo2017deepfm}. It combines factorization machine and DNN.
\item \textbf{AutoInt} \cite{song2019autoint}. It consists of a multi-head self-attentive network with residual connections and DNN.
\item \textbf{FedSplitNN} \cite{he2023hybrid}. A classical VFL method using aligned data.
\item \textbf{FedCTR} \cite{wu2022fedctr}. It exploits user features on other platforms to improve CTR prediction on the advertising platform.
\item \textbf{SS-VFL} \cite{castiglia2022self}. Self-Supervised Vertical Federated Learning. Step 1: each party uses its local data to pre-train network parameters using self-supervised learning. Step 2: aligned data are used to train a downstream prediction task.
\item \textbf{FedHSSL} \cite{he2023hybrid}. Federated Hybrid Self-Supervised Learning. Step 1: cross-party self-supervised learning \cite{grill2020bootstrap,chen2021exploring} using aligned data. Step 2: cross-party-guided local self-supervised learning using local data. Step 3: partial model aggregation.
\item \textbf{FedUD}. The proposed federated learning with unaligned data method in this paper.
\end{itemize}

Among these methods, DNN, Wide\&Deep, DeepFM and AutoInt are local methods which use only the host party's data to train a local model.
The others are VFL methods. In particular, FedSplitNN and FedCTR use only aligned data across parties. SS-VFL, FedHSSL and FedUD use both aligned and unaligned data. All the above methods are based on standard CTR features per sample.

\begin{table}[!t]
\setlength{\tabcolsep}{2pt}
\renewcommand{\arraystretch}{1.1}
\caption{Statistics of experimental datasets.}
\vskip -11pt
\label{tab_stat}
\centering
\begin{tabular}{|l|c|c|c|c|c|c|}
\hline
\textbf{Dataset} & \textbf{\# Fields} & \textbf{\# Train} & \textbf{\# Val} & \textbf{\# Test} & \textbf{\# Show} & \textbf{\# Click} \\
\hline
Avazu & 22 & 32.4M & 3.83M & 4.22M & 40.43M & 6.86M \\
\hline
Industrial & 34 & 439.3M & 62.5M & 61.7M & 563.5M & 209.8M \\
\hline
\end{tabular}
\vskip -12pt
\end{table}

\subsection{Settings}
\textbf{Parameters.} We set the embedding dimension as 10, the layer dimensions in the top model as \{256, 128\} and those in other models as \{512, 256, 128\}.

\textbf{Evaluation Metric.} We use AUC and LogLoss as the evaluation metrics for CTR prediction.

\begin{table*}[!t]
\setlength{\tabcolsep}{1.5pt}
\renewcommand{\arraystretch}{1.08}
\caption{Test AUCs on experimental datasets. ``overall'', ``aligned'' and ``unaligned'' mean AUC is computed on all the test data, only aligned test data (which have the guest party's features) and only unaligned test data (which do not have the guest party's features) respectively. The best result is in bold font. The second best result is underlined. A \textnormal{small} improvement in AUC (e.g., 0.0020) can lead to a \textnormal{significant} increase in online CTR (e.g., 3\%) \cite{cheng2016wide}. * indicates the statistical significance for $p \leq 0.01$ compared with the second best result over paired t-test.}
\vskip -10pt
\label{tab_auc}
\centering
\begin{tabular}{|l|c c|c c|c c|c c|c c|c c|}
\hline
 & \multicolumn{6}{|c|}{\textbf{Avazu}} & \multicolumn{6}{|c|}{\textbf{Industrial}} \\
\hline
 & \multicolumn{2}{|c|}{overall} & \multicolumn{2}{|c|}{aligned} & \multicolumn{2}{|c|}{unaligned}
 & \multicolumn{2}{|c|}{overall} & \multicolumn{2}{|c|}{aligned} & \multicolumn{2}{|c|}{unaligned} \\
 \hline
& AUC $\uparrow$ & LogLoss $\downarrow$ & AUC$\uparrow$ & LogLoss$\downarrow$ & AUC$\uparrow$ & LogLoss$\downarrow$
& AUC$\uparrow$ & LogLoss$\downarrow$  & AUC$\uparrow$ & LogLoss$\downarrow$ & AUC$\uparrow$ & LogLoss$\downarrow$ \\
\hline
DNN                 & 0.7186 & 0.4140 & 0.7203 & 0.4204 & 0.6997 & 0.3695 & 0.7996 & 0.5066 & 0.8032 & 0.4977 & 0.7909 & 0.5253 \\
Wide\&Deep & 0.7178 & 0.4154 & 0.7191 & 0.4211 & 0.6994 & 0.3696 & 0.7997 & 0.5066 & 0.8034 & 0.4977 & 0.7909 & 0.5253 \\
DeepFM          & 0.7185 & 0.4142 & 0.7202 & 0.4204 & 0.6995 & 0.3696 & 0.7997 & 0.5067 & 0.8033 & 0.4977 & 0.7910 & 0.5253 \\
AutoInt            & 0.7196 & 0.4111 & 0.7212 & 0.4187 & \uwave{0.7003} & \uwave{0.3694} & 0.7997 & 0.5066 & 0.8033 & 0.4977 & \uwave{0.7912} & \uwave{0.5252} \\
\hline
FedSplitNN    & 0.7283 & 0.4074 & 0.7330 & 0.4120 & 0.6932 & 0.3809 & 0.7990 & 0.5081 & 0.8102 & 0.4911 & 0.7761 & 0.5437 \\
FedCTR           & 0.7294  & 0.4068 & 0.7346 & 0.4112 & 0.6935 & 0.3807 & 0.7998 & 0.5068 & 0.8099 & 0.4918 & 0.7786 & 0.5405 \\
SS-VFL             & 0.7298 & 0.4065 & 0.7351 & 0.4109 & 0.6948 & 0.3797 & \uwave{0.8001} & \uwave{0.5063} & \uwave{0.8106} & \uwave{0.4908} & 0.7778 & 0.5412 \\
FedHSSL        &\uwave{0.7302}  & \uwave{0.4063} & \uwave{0.7358} & \uwave{0.4103} & 0.6941 & 0.3805 & 0.7999 & 0.5065 & 0.8105 & 0.4909 & 0.7771 & 0.5419 \\
\hline
FedUD           & \textbf{0.7355}$^*$ & \textbf{0.4037}$^*$ & \textbf{0.7370}$^*$ & \textbf{0.4093}$^*$ & \textbf{0.7083}$^*$ & \textbf{0.3652}$^*$
& \textbf{0.8057}$^*$ & \textbf{0.5026}$^*$ & \textbf{0.8121}$^*$ & \textbf{0.4899}$^*$ & \textbf{0.7952}$^*$ &\textbf{0.5224}$^*$ \\
\hline
\end{tabular}
\vskip -8pt
\end{table*}

\subsection{Experimental Results}

\subsubsection{\textbf{Effectiveness}}
In Table \ref{tab_auc}, ``overall'', ``aligned'' and ``unaligned'' mean AUC/LogLoss is computed on all the test data, only aligned test data (which have the guest party's features) and only unaligned test data (which do not have the guest party's features) respectively. It is observed that most federated methods (e.g., FedHSSL) perform better than local methods (e.g., AutoInt) on aligned test data because of the inclusion of the guest party' features. However, most federated methods perform worse than local methods on unaligned test data because of the absence of the guest party's features. Local methods do not use the guest party's features at all. Although SS-VFL and FedHSSL also exploit unaligned data, they only use such data for self-supervised learning, and thus they also do not perform well on unaligned test data. Differently, FedUD explicitly transfers knowledge from aligned data to unaligned data and exhibits the best performance on both aligned and unaligned data.
It is observed that the AUC improvement of FedUD over DNN (a local method) on aligned data is higher than that on unaligned data. It is because aligned data have \emph{real} guest party's features but unaligned data have \emph{inferred} guest party's representations which are more noisy.

\subsubsection{\textbf{Effect of the Guest Party's Feature Slots}}

\begin{figure}[!t]
\centering
\subfigure[Avazu]{\includegraphics[width=0.235\textwidth, trim = 0 0 0 0, clip]{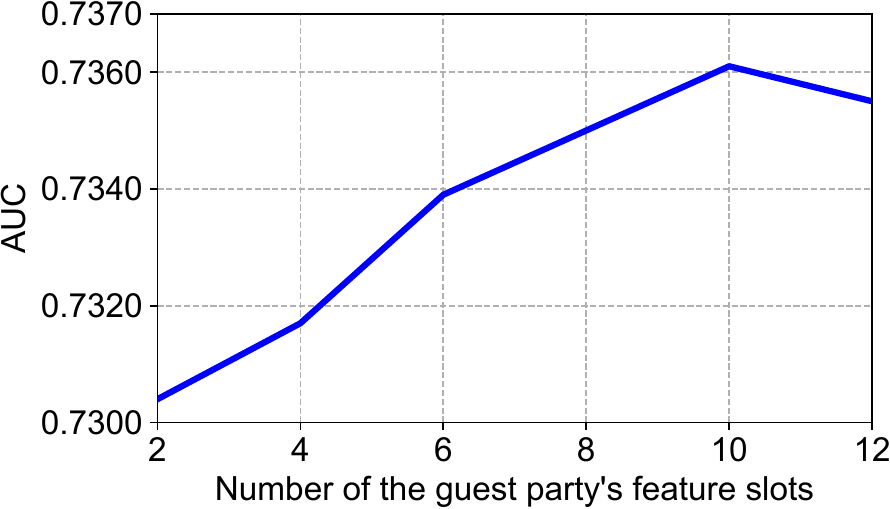}}
\subfigure[Industrial]{\includegraphics[width=0.235\textwidth, trim = 0 0 0 0, clip]{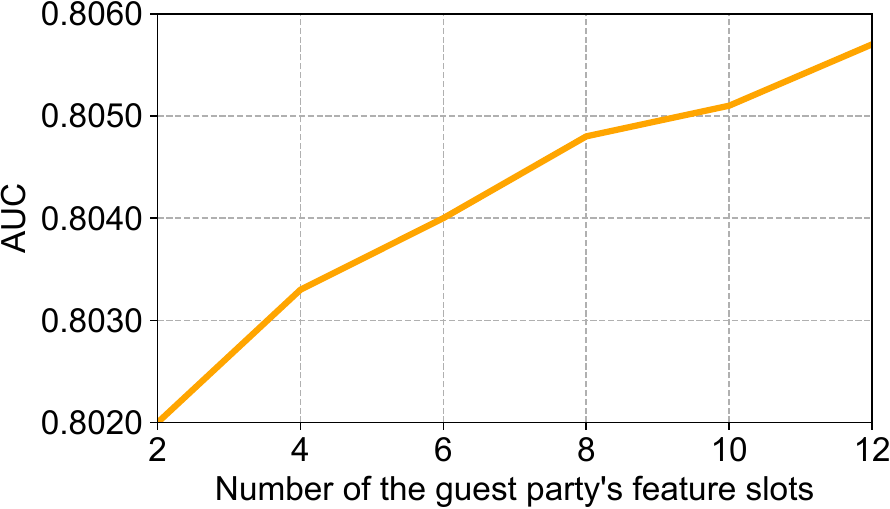}}
\vskip -12pt
\caption{Overall test AUC vs. the number of the guest party's feature slots. (a) Avazu dataset. (b) Industrial dataset.}
\vskip -10pt
\label{fig_slot}
\end{figure}

Figure \ref{fig_slot} shows the overall AUCs vs. the number of the guest party's feature slots. AUCs on aligned / unaligned test data show similar trends. Due to space limitation, we omit these figures. It is observed that more guest party's feature slots generally provide useful additional information and lead to improved performance. But adding more feature slots may also include noise and possibly lead to flatten or even slightly degraded performance.

\subsubsection{\textbf{Effect of the Host Party's Unaligned Samples}}
\begin{figure}[!t]
\centering
\subfigure[Avazu]{\includegraphics[width=0.235\textwidth, trim = 0 0 0 0, clip]{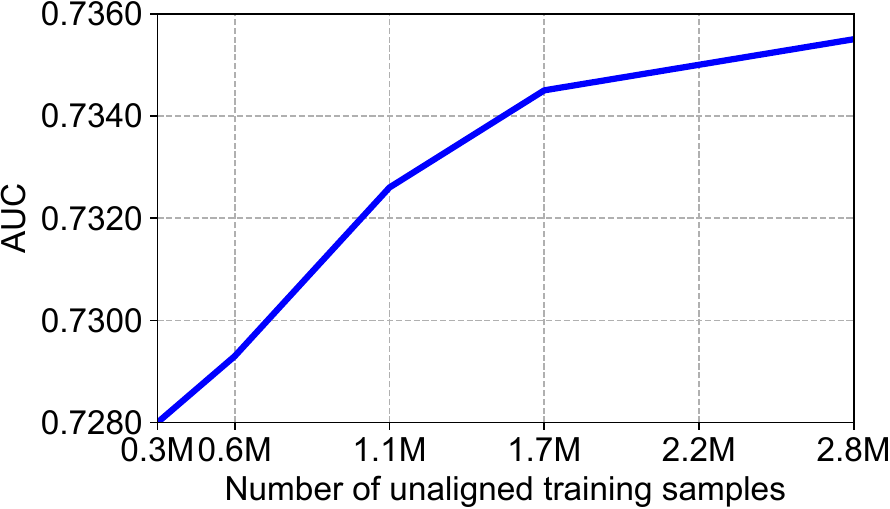}}
\subfigure[Industrial]{\includegraphics[width=0.235\textwidth, trim = 0 0 0 0, clip]{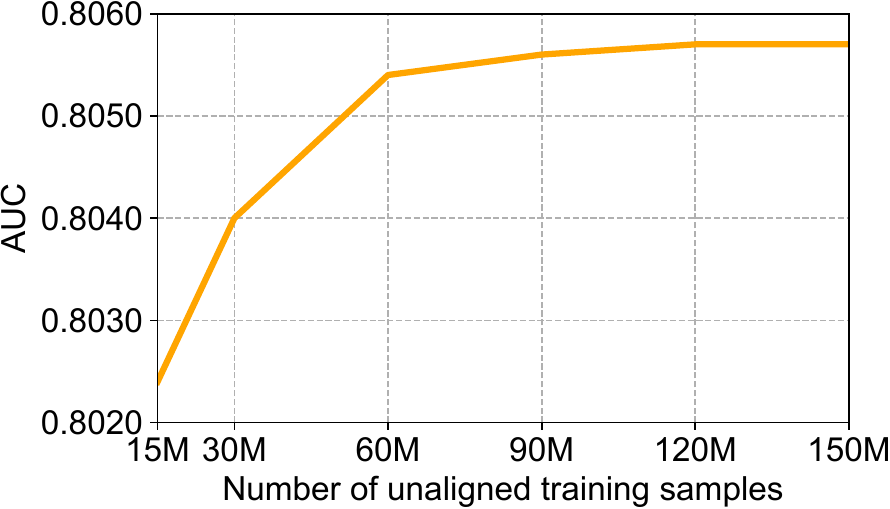}}
\vskip -12pt
\caption{Overall test AUC vs. the number of unaligned training samples. (a) Avazu dataset. (b) Industrial dataset.}
\vskip -10pt
\label{fig_un}
\end{figure}

In this experiment, we use all the aligned samples but different numbers of unaligned samples for training.
Figure \ref{fig_un} shows the overall AUCs vs. the number of the host party's unaligned samples. The rightmost point on the x-axis denotes the maximum number of unaligned samples in the training set. It is observed that more unaligned samples generally lead to improved prediction performance. But after certain amount, the improvement becomes less obvious.

\section{Related Work}
\textbf{CTR prediction.}
CTR prediction has attracted lots of attention in recent years. Methods range from shallow models such as LR \cite{richardson2007predicting}, FM \cite{rendle2010factorization},  Field-weighted FM \cite{pan2018field} to deep models such as Wide \& Deep \cite{cheng2016wide}, DeepFM \cite{guo2017deepfm}, xDeepFM \cite{lian2018xdeepfm}, AutoInt \cite{song2019autoint} and DIL \cite{zhang2023reformulating}.
These methods are based on standard CTR features per sample. Some other methods exploit auxiliary information. DIN \cite{zhou2018deep} and DIEN \cite{zhou2019deep} consider user historical click behaviors. BERT4CTR \cite{wang2023bert4ctr} considers pre-trained language model information.
All these methods are centralized, which process all the data in a central server.

\textbf{Vertical federated learning (VFL).}
VFL provides a feasible solution for cross-platform federated CTR prediction. It is able to keep the raw data locally and learn a collaborative model in a privacy-preserving way \cite{zhang2021survey,yang2019federated,kang2022privacy}. However, FedSplitNN \cite{he2023hybrid} and FedCTR \cite{wu2022fedctr} only utilize aligned data with common keys across parties, which strongly restricts their application scope. SS-VFL \cite{castiglia2022self} and FedHSSL \cite{he2023hybrid} exploit both aligned and unaligned data, but they only use unaligned data for local self-supervised learning. Differently, FedUD proposed in this paper explicitly transfers knowledge learned from cross-party aligned data to unaligned data. FedUD is thus able to benefit both aligned and unaligned inference data.

\section{Conclusion}
In this paper, we propose FedUD for federated CTR prediction. FedUD is able to transfer the knowledge from aligned data across parties to enrich the representations of unaligned data such that both aligned and unaligned data can contribute to federated model training. It also benefits both aligned and unaligned data during inference. Experimental results demonstrate the effectiveness of FedUD for federated CTR prediction.

\bibliographystyle{ACM-Reference-Format}
\balance
\bibliography{ref}

\end{document}